\documentclass[twocolumn, tighten, dvipsnames, floatfix]{aastex63}

\usepackage[]{xcolor}
\usepackage{booktabs}
\usepackage{physics}
\usepackage{wasysym}

\usepackage{longtable}
\usepackage{threeparttablex} 

\usepackage{array}
\usepackage{amsmath}
\usepackage{changepage}

\usepackage{lineno, ulem} 

\usepackage[flushmargin]{footmisc}
\usepackage{afterpage}
\usepackage{subfigure}

\newcommand{\lzuni}{kpc km/s}

\def\arcsec{\hbox{$^{\hbox{\rlap{\hbox{\lower4pt\hbox{$\,\prime\prime$}}
          }}}$} \ }

\def\arcmin{\hbox{$^{\hbox{\rlap{\hbox{\lower4pt\hbox{$\;\prime$}}
          }\hbox{$\frown$}}}$}}

\usepackage{lipsum}
\usepackage{float}

\shorttitle{The Unmixed Debris of Gaia-Sausage/Enceladus}
\shortauthors{Perottoni et al.}

\begin{document}

\title{The Unmixed Debris of Gaia-Sausage/Enceladus in the Form of \\ a Pair of Halo Stellar Overdensities}


\correspondingauthor{Hélio D. Perottoni}
\email{hperottoni@gmail.com}

\author[0000-0002-0537-4146]{H\'elio D. Perottoni}
\affil{Universidade de S\~ao Paulo, Instituto de Astronomia, Geof\'isica e Ci\^encias Atmosf\'ericas, Departamento de Astronomia, \\ SP 05508-090, S\~ao Paulo, Brazil}

\author[0000-0002-9269-8287]{Guilherme Limberg}
\affil{Universidade de S\~ao Paulo, Instituto de Astronomia, Geof\'isica e Ci\^encias Atmosf\'ericas, Departamento de Astronomia, \\ SP 05508-090, S\~ao Paulo, Brazil}

\author[0000-0002-7662-5475]{Jo\~ao A. S. Amarante}\altaffiliation{Visiting Fellow at UCLan}
\affiliation{Institut de Ciències del Cosmos (ICCUB), Universitat de Barcelona (IEEC-UB), Martí i Franquès 1, E08028 Barcelona, Spain}
\affil{Jeremiah Horrocks Institute, University of Central Lancashire, Preston, PR1 2HE, UK}

\author[0000-0001-7479-5756]{Silvia Rossi}
\affil{Universidade de S\~ao Paulo, Instituto de Astronomia, Geof\'isica e Ci\^encias Atmosf\'ericas, Departamento de Astronomia, \\ SP 05508-090, S\~ao Paulo, Brazil}

\author[0000-0001-9209-7599]{Anna B. A. Queiroz}
\affil{Leibniz-Institut f$\ddot{u}$r Astrophysik Potsdam (AIP), An der Sternwarte 16, D-14482 Potsdam, Germany}
\affil{Institut f\"{u}r Physik und Astronomie, Universit\"{a}t Potsdam, Haus 28 Karl-Liebknecht-Str. 24/25, D-14476 Golm (Potsdam), Germany}

\author[0000-0002-7529-1442]{Rafael M. Santucci}
\affiliation{Universidade Federal de Goi\'as, Instituto de Estudos Socioambientais, Planet\'ario, Goi\^ania, GO 74055-140, Brazil}
\affiliation{Universidade Federal de Goi\'as, Campus Samambaia, Instituto de F\'isica, Goi\^ania, GO 74001-970, Brazil}

\author[0000-0002-5974-3998]{Angeles P\'erez-Villegas}
\affil{Instituto de Astronom\'ia, Universidad Nacional Aut\'onoma de M\'exico, Apartado Postal 106, C. P. 22800, Ensenada, B. C., M\'exico}

\author[0000-0003-1269-7282]{Cristina Chiappini}
\affil{Leibniz-Institut f$\ddot{u}$r Astrophysik Potsdam (AIP), An der Sternwarte 16, D-14482 Potsdam, Germany}
\affil{Laborat\'orio Interinstitucional de e-Astronomia - LIneA, RJ 20921-400, Rio de Janeiro, Brazil}

\begin{abstract}
 
In the first billion years after its formation, the Galaxy underwent several mergers with dwarf satellites of various masses. The debris of Gaia-Sausage/Enceladus (GSE), the galaxy responsible for the last significant merger of the Milky Way, dominates the inner halo and has been suggested to be the progenitor of both the Hercules-Aquila Cloud (HAC) and Virgo Overdensity (VOD). We combine SEGUE, APOGEE, \textit{Gaia}, and \texttt{StarHorse} distances to  characterize the chemodynamical properties and verify the link between HAC, VOD, and GSE. We find that the orbital eccentricity distributions of the stellar overdensities and GSE are comparable. We also find that they have similar, strongly peaked, metallicity distribution functions, reinforcing the hypothesis of common origin. Furthermore, we show that HAC and VOD are indistinguishable from the prototypical GSE population within all chemical-abundance spaces analyzed. All these evidences combined provide a clear demonstration that the GSE merger is the main progenitor of the stellar populations found within these halo overdensities.

\end{abstract}


\section{Introduction}
\label{sec:intro}

\setcounter{footnote}{4}

The concordance cosmological model of galaxy formation predicts that the halos of massive galaxies, such as the Milky Way, are formed via the accretion of many low-mass satellites \citep{Springel2006}. While the star-by-star characterization of ancient accretion events in other galaxies is still unfeasible, wide-area photometric and spectroscopy surveys have shown the complexity of the Milky Way's stellar halo (or simply ``halo"; \citealt{Ivezic2012,Belokurov2013_review}) which includes, e.g., several stellar overdensities \citep{Newbergcarlin2016Book}. These overdensities are identified as stellar count excesses in given Galactic regions when compared to other homogeneous/smooth halo fields. \par
The stellar halo overdensities can form via the buildup of tidal debris, from one or more satellite galaxies, whose stars have highly eccentric orbits that accumulate at the apocenter \citep{Newbergcarlin2016Book,Li2016EriPho,Deason2018}, a scenario that has been explored with pure $N$-body models \citep{Johnston2008,Helmi2011, Naidu2021simulations}.
Moreover, the accretion of a single dwarf galaxy can create more than a single stellar overdensity, challenging the association between distinct substructures and their progenitors \citep[e.g.,][]{Johnston2012}.
This hypothesis has gained recent attention with the confirmation of the last significant merger (mass ratio $\gtrsim 1{:}4$) experienced by the Milky Way with a dwarf galaxy named Gaia-Sausage/Enceladus (GSE; \citealt{belokurov2018, helmi2018}). The GSE merger happened $\sim$10 Gyr ago \citep{gallart2019,Bonaca2020,montalban2021} and is likely responsible for many features observed in our Galaxy \citep{Deason2018,DiMatteo2019,Iorio2019,belokurov2020}. \par
Currently, there are two overdensities that have been tentatively linked to GSE, the Hercules-Aquila Cloud (HAC) and the Virgo Overdensity (VOD). HAC was discovered as a diffuse substructure of main-sequence turnoff stars located toward the Galactic center \citep{Belokurov2007HAC}. It extends in heliocentric distances ($d_\odot$) from 10 to 20 kpc,  from 25$^{\circ} < l < 60^{\circ}$ and $-$40$^{\circ} < b < 40^{\circ}$  \citep{Belokurov2007HAC,Watkins2009,Sesar2010,Simion2014}.  \citet{Watkins2009} and \citet{Sesar2010}, based on photometric metallicities of RR Lyrae stars, measured $\langle \rm[Fe/H] \rangle = -1.42$ and $-1.50$, respectively. Recently, \citet{Naidu2021simulations}, using spectroscopic data from the Hectochelle in the Halo at High Resolution survey (H3; \citealt{Conroy2019A}), measured a median metallicity of $\rm[Fe/H] = -1.20$. \par 

VOD was first identified with RR Lyrae and main-sequence turnoff stars as an excess in the stellar halo \citep{Vivas2001,Newberg2002}. Later, these were associated as part of a larger overdensity located in the direction of the Virgo constellation \citep{Juric2008}.
Its main stellar excess is located between 270$^{\circ} < l < 330^{\circ}$ and 50$^{\circ} < b < 75^{\circ}$, and possibly extends over a larger area on the sky ($\sim$3000 deg$^{2}$; \citealt{Juric2008,Bonaca2012A}).
Its $\langle \rm[Fe/H] \rangle$ varies from $-1.1$ to $-1.95$ depending on the stellar sample \citep{Vivas2008,Carlin2012,Naidu2021simulations}, and the stars associated with VOD are typically on high-eccentricity orbits ($\langle e \rangle \sim 0.8$; \citealt{Carlin2012,Simion2019}). \par
%
\citet{Simion2019} explored the association between the GSE merger and both VOD and HAC using astrometric data from the Gaia space mission \citep{GaiaMission}. They showed that these overdensities share similar dynamical properties, such as the orbital energy, apocentric distance, and eccentricity, with GSE stars and interpreted it as an evidence that HAC and VOD are unmixed debris of GSE. This is also suggested by the apparent orbital pileup of local halo stars in the regions associated with these overdensities \citep[][]{Balbinot2021} and with pure $N$-body models mimicking the GSE merger \citep{Naidu2021simulations}. \par

Due to their relative large distances from the Sun, it is difficult to obtain, for a large number of stars, accurate abundances and reliable distance measurements for members of these substructures. We overcome this challenge by taking advantage of the \texttt{StarHorse} code \citep{Santiago2016starhorse, Queiroz2018}, which is as a Bayesian isochrone-fitting method able to derive distances, extinctions, and other stellar parameters based on observed spectroscopic, photometric, and astrometric data. Crucially, these precise \texttt{StarHorse} distance estimates allows to confidently select members of these overdensities in order to conduct a chemical comparison between HAC/VOD and nearby GSE stars. This exercise is necessary to confirm (or reject) the link between the overdensities and the main accretion event of our Galaxy. Our goal is to verify whether \textit{i}) they have solely an accreted origin, \textit{ii}) they have similar chemodynamical properties among them, and \textit{iii}) they are (in)distinguishable from GSE.

\begin{figure*}[ht!]%
\includegraphics[width=1.5\columnwidth, trim = 0cm -2cm 0cm 0cm]{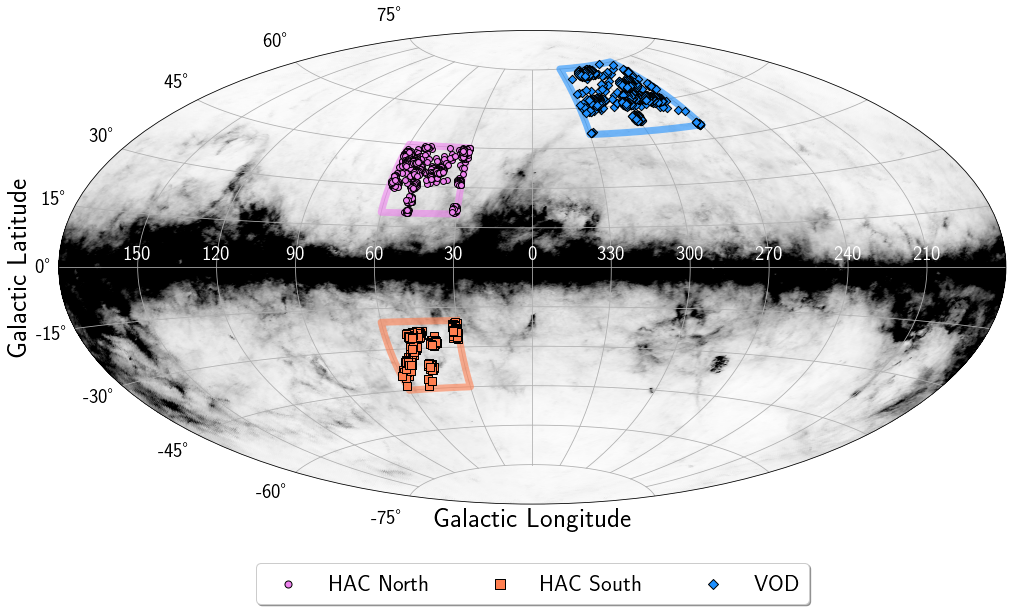}
\includegraphics[width=0.6\columnwidth]{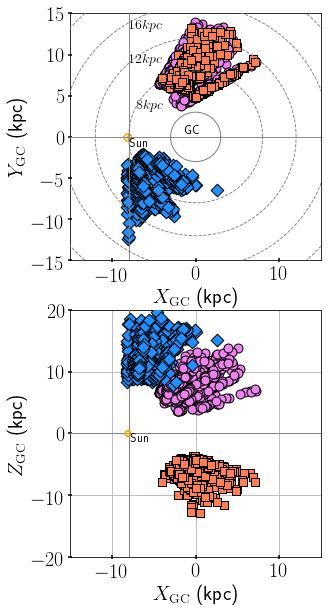}
\caption{Spatial projection in Galactic coordinates of the SEGUE sample selection of the stellar overdensities (left). Distribution of the sample of stars from SEGUE in the $X_{\rm GC} - Y_{\rm GC} $ (top right) and  $X_{\rm GC}- Z_{\rm GC}$ projections (bottom right). The HAC-S, HAC-N, and VOD are represented by orange squares, violet circles, and blue diamonds, respectively. The rings are concentric with the center of the Galaxy.}
\label{fig:projection}
\end{figure*}

In this spirit, we carry out an in-depth chemodynamical characterization of HAC and VOD with data from Apache Point Observatory Galactic Evolution Experiment (APOGEE; \citealt{apogee2017}) and Sloan Extension for Galactic Understanding and Exploration (SEGUE; \citealt{yanny2009}). 
With the new data, we are now able to demonstrate that the majority of the stars from the stellar overdensities are consistent with the prototypical population of GSE defined within 5 kpc around the Sun. 

\section{Data} 
\label{sec:data}

\subsection{SEGUE, APOGEE, and {\it Gaia}} 

We keep only those stars from SEGUE \citep{SEGUE2022} with $S/N>20$ pixel$^{-1}$. We limit our SEGUE sample to $\rm[Fe/H] < -0.5$ to decrease contamination from disk stars. We also select stars within $4500 < T_{\rm eff}/{\rm K} < 6500$, which ensures the stellar parameters are accurately determined and avoids large uncertainties. The abundance data from SEGUE are less precise (uncertainties in [Fe/H] $\geq 0.1$), but with the advantage of observing toward fainter-magnitude stars, providing a larger sample for each overdensity, which allows a robust determination of the orbital eccentricity and metallicity distributions. To obtain reliable abundances for a larger set of elements, we expand our analysis using APOGEE data. \par

In order to obtain stars with high-quality elemental abundances, we select only those sources from APOGEE data release 17 (DR17; \citealt{APOGEEdr17}) catalog with good spectroscopic flags ({\tt ASPCAPFLAG == 0}), good synthetic spectral fitting ({\tt ASPCAP\_CHI2 $<$ 25}), good estimates of [Fe/H], [Mg/Fe], [Al/Fe], [Mn/Fe], [Ni/Fe], [C/Fe], [N/Fe], and [O/Fe] (i.e., flagged == 0), $S/N > 50$ pixel$^{-1}$, and we remove stars with suspect radial velocities ({\tt STARFLAG == SUSPECT\_RV\_COMBINATION}). Lastly, we apply additional cuts ($4000 < T_{\rm eff}/{\rm K} < 6000$, $\log g < 3$) in order to select only giant stars. \par
\par
We combine (1.5'' search radius) the aforementioned samples with Gaia Early Data Release 3 \citep{GaiaEDR3Summary} to obtain the absolute proper motions and their uncertainties. To ensure good astrometric solutions, we impose renormalized unit weight errors within the recommended range ($\texttt{RUWE} \leq 1.4$; \citealt{Lindegren2020a}). Besides that, we remove stars with $\texttt{parallax} < -5$ and with low fidelity ($\texttt{fidelity\_v2} < 0.5$; see \citealt{Rybizki2022fidelity}). 
The radial velocities are from the spectroscopic surveys, and the distances were obtained with the \texttt{StarHorse} code, specifically from the new releases where several large-scale spectroscopic surveys are matched to Gaia EDR3 data. For APOGEE DR17, distance estimates are publicly available\footnote{\url{https://www.sdss.org/dr16/data_access/value-added-catalogs/?vac_id=apogee-dr17-starhorse-distances,-extinctions,-and-stellar-parameters}.} and, for SEGUE, these will be made available alongside a forthcoming paper (Queiroz et al., in preparation). We selected only stars with fractional (Gaussian) uncertainties of their nominal distance values ${<}20\%$.

We distinguish HAC into two substructures HAC-South (HAC-S) and HAC-North (HAC-N). Under the assumption that both regions have the same  origin \citep{Iorio2019,Simion2019, Naidu2021simulations}, they should have similar chemodynamical properties. We select stars from HAC-S and HAC-N with the following criteria: 30$^{\circ} < l < 60^{\circ}$, $-$45$^{\circ} < b < -20^{\circ}$ and $20^{\circ} < b < 45^{\circ}$, respectively.
For VOD, our selection is based on the following Galactic coordinate cuts: 270$^{\circ} < l < 330^{\circ}$ and 50$^{\circ} < b < 75^{\circ}$. For both overdensities, we consider $d_\odot$ between 10 and 20 kpc, with a mean fractional uncertainty of $\sim$13\% in this range. These delineated criteria were applied to both SEGUE and APOGEE samples. Figure \ref{fig:projection} shows the resulting spatial distribution of stars selected from HAC and VOD in SEGUE.

\subsection{Orbital Parameters} 
\label{sec:orbit_para}

To obtain the orbital parameters, we employ the axisymmetric Galactic potential of \citet{mcmillan2017}. The adopted distance from the Sun to the Galactic center is 8.2\,kpc \citep{BlandHawthorn2016}, the circular velocity at this position is $v_{\rm circ} = 232.8$\,km\,s$^{-1}$, and the peculiar motion of the Sun with respect to the local circular orbit is $(U,V,W)_\odot = (11.10, 12.24, 7.25)$\,km\,s$^{-1}$ \citep{schon2010}. We integrated the orbits forward in time for 10 Gyr. For each star, we construct a set of 100 initial conditions using a Monte Carlo technique taking into account the observational uncertainties in $d_\odot$, proper motions, and the line-of-sight velocities. The final dynamical parameters are taken as the medians of the derived distributions and associated uncertainties are the 16$^{\rm th}$ and 84$^{\rm th}$ percentiles.\par

\begin{figure*}[ht!]
\includegraphics[width=18.0cm]{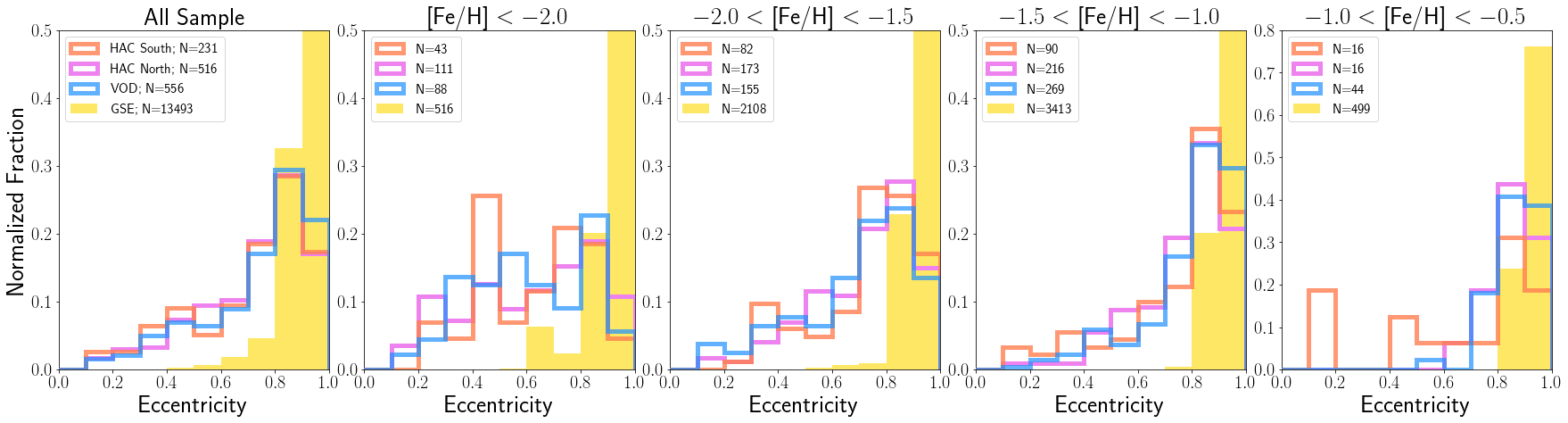}
\caption{Orbital eccentricity distribution of SEGUE stars from HAC-S, HAC-N, VOD, and GSE in orange, pink, blue, and yellow, respectively. The histogram shows the fractions of stars in each bin, and the number of stars ($N$) in each panel is also provided. The first panel shows the eccentricity distribution of the full sample and each of the other panels corresponds to a slice in metallicity, with [Fe/H] increasing from left to right.}
\label{fig:segue_ecc_hist}
\end{figure*}

%
In order to compare whether the stellar overdensities share similar dynamical properties with GSE, we use a chemodynamical criteria (\citealt{Limberg2022}; see their equation 1)\footnote{We note that, since SEGUE does not provide abundances other than [Fe/H] and [$\alpha$/Fe], we do not apply the chemical portion of these author's criteria (which demand Mg, Mn, and Al).} to identify nearby---within 5 kpc from the Sun---GSE stars, which is designed to yield maximum purity. The final SEGUE/APOGEE samples of HAC-S, HAC-N, and VOD contain 231/11, 516/21, and 556/22 stars, respectively. 

\section{Analysis and Discussion}\label{sec:ana}

\subsection{Eccentricity and Metallicity Distributions}
\label{sec:segue_analy}

We take advantage of the large number of the HAC-S, HAC-N, and VOD stars found in SEGUE to characterize the orbital eccentricity and metallicity distributions of these stellar overdensities with the purpose of testing their connection with GSE. We explore their distributions of orbital eccentricity in Figure \ref{fig:segue_ecc_hist}. The first panel shows the full sample, and overall, the overdensities have a similar eccentricity distributions; they are dominated (${\sim}65\%$) by stars with $e > 0.7$, similar to GSE \citep{belokurov2018,Myeong2018,Limberg2021}, but have an extended tail toward lower eccentricity values. \par

We also show the eccentricity distribution at different metallicity intervals (second to fifth panels of Figure \ref{fig:segue_ecc_hist}). The overdensities have similar distributions across all the metallicity intervals but differ from the GSE. For instance, in the very metal-poor regime (VMP; $\rm[Fe/H] < -2.0$), the overdensities have roughly equal amounts of stars in each eccentricity bin, whereas the GSE continues to be dominated by stars with $e > 0.7$.  Toward higher metallicities ($-1.5 < \rm[Fe/H] < -1.0$), where the peak of the GSE metallicity distribution function (MDF) is located (see \citealt{Limberg2022} and references therein), the overdensities are dominated by stars with $e > 0.7$ and, differently from GSE, show an extended tail toward lower eccentricities. 

\par

In addition, Figure \ref{fig:segue_ecc_hist} shows that the contribution of low-eccentricity stars is higher toward the low-metallicity regime ($\rm[Fe/H] < -1.5$). The absence of the low-eccentricity tail in the GSE sample could be a bias due to our selection criterion, which favors stars with vertical component of the angular momentum ($L_z$) $\sim 0$. On the other hand, simulations mimicking GSE-like mergers \citep{Naidu2021simulations,amarante+2022} show that the vast majority of the accreted stars end up with $e > 0.7$. Indeed, models that take into account star formation and a self-consistent chemical enrichment show that the accreted stars on relatively lower eccentricities are more metal-poor compared to those on higher eccentricities (see figure 6 in \citealt{amarante+2022}). Only a small fraction (${\sim}1{-}8\%$) of the accreted stars end up on less eccentric orbits ($e < 0.7$) and these are typically more metal-poor (by ${\sim}0.2{-}0.7$ dex) than those with $e>0.7$. Therefore, part of the metal-poor stars on low-eccentricity orbits found in SEGUE could be the metal-poor tail of GSE, as expected by the aforementioned idealized GSE-like merger models. We note that the metal-poor stars in our HAC/VOD sample have $-1500 \lesssim L_z \lesssim 1500$ \lzuni\, and have roughly the same fraction of prograde and retrograde orbits. Thus, other mildly eccentric ($e \sim 0.5$) exclusively retrograde ($L_z \lesssim -1500$ \lzuni) halo structures can be discarded as responsible for the low-metallicity and low-eccentricity component identified here.

\par
%
\begin{figure*}[ht!]
\includegraphics[width=18.0cm]{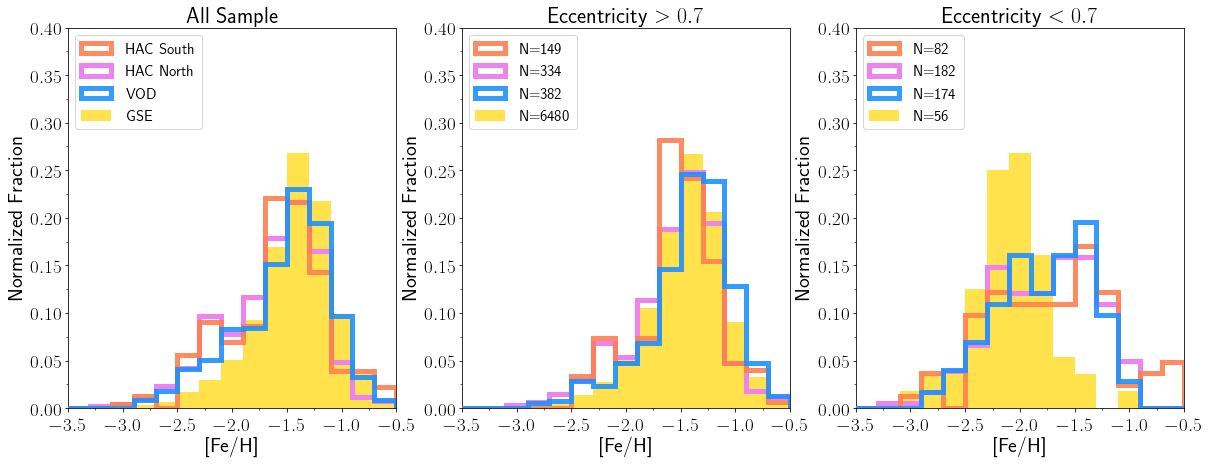}
\caption{The MDFs of the SEGUE stars from HAC-S, HAC-N, VOD, and GSE. 
The MDFs of the full sample (left); stars with eccentricity higher than 0.7 (middle) and lower than 0.7 (right) are presented  with the same color scheme of Figure \ref{fig:segue_ecc_hist}.} 
\label{fig:segue_hist_metal}
\end{figure*}

%

The presence of the low-eccentricity stars in both HAC-S, HAC-N, and VOD may also be interpreted as the existence of other merger(s) event(s) that contributed less significantly to the formation of these overdensities. We verified that the upper limit of contamination by stars from the Sagittarius stream in both overdensities is ${<}5\%$ following the criteria of \citet{naidu2020} \citep[see also][]{Penarrubia2021sgr}. Another potential contributor to the low-eccentricity regime  might be LMS-1/Wukong \citep{yuan2020,naidu2020,Horta2022}, which contains mostly VMP stars with $e < 0.7$
. This structure is also composed of stars on polar orbits ($J_z > J_r$) that spatially overlap with VOD and HAC \citep{yuan2020}. We note, however, that LMS-1 members are exclusively on prograde orbits, which is not the case for the VMP low-eccentricity stars in our HAC/VOD sample.

\begin{figure*}[ht!]
\includegraphics[width=18.0cm]{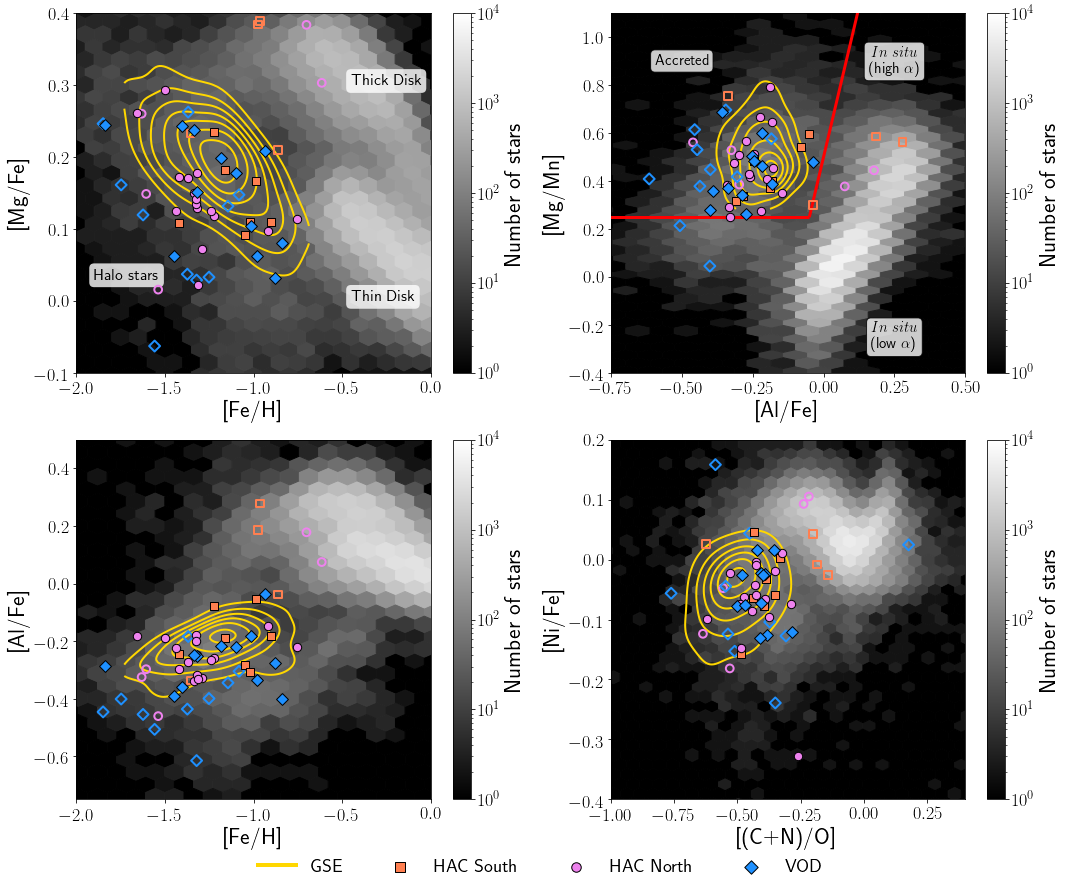}
\caption{Chemical abundances of HAC-S, HAC-N, and VOD stars in the [Mg/Fe]--[Fe/H] (top left), [Mg/Mn]--[Al/Fe] (top right), [Al/Fe]--[Fe/H] (bottom left), and [Ni/Fe]--[(C+N)/O] (bottom right). 
The empty symbols show stars from the stellar overdensities that either do not have the typical elemental abundances of ancient accreted populations (region in the top right panel delineated by red lines \citealt{Hawkins2015, das2020}) or have low-eccentricity orbits ($e < 0.7$). The yellow contours indicate the chemical patterns of the GSE stars. The background 2D histogram shows the distribution of the APOGEE stars.} %
\label{fig:apogee_abundance}
\end{figure*}

Finally, \textit{in situ} stellar populations could contribute to the stellar overdensities, such as the Splash \citep{belokurov2020} and Aurora \citep{Belokurov2022aurora}. The Splash consists of metal-rich ($\rm[Fe/H] > -0.7$) stars on mildly eccentric orbits ($e > 0.5$), while the majority of HAC and VOD low-eccentricity stars have $\rm[Fe/H] < -1.5$. Aurora is mainly characterized by its thick-disk-like chemistry, i.e., high [$\alpha$/Fe] at the same [Fe/H] in comparison to the Galactic halo/GSE. Moreover, Aurora should be identifiable at metallicities of $-1.5 < \rm[Fe/H] < -1.0$, which is not the case for the low-eccentricity populations identified within HAC/VOD. Therefore, given the present data, we conclude that is is unlikely that an \textit{in situ} contribution could be the major non-GSE contaminant to the stellar overdensities.

\begin{figure*}[ht!]
\includegraphics[width=18.0cm]{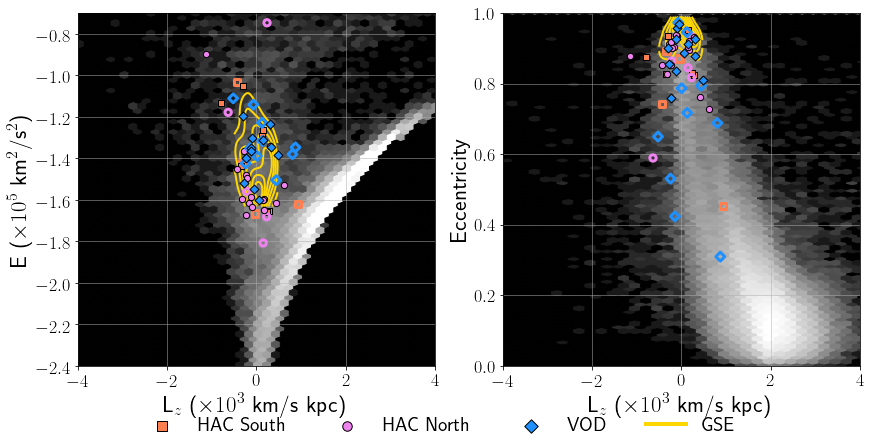}
\caption{Orbital parameters of HAC-S, HAC-N, and VOD stars in the $E$--$L_z$ (left) and $e$--$L_{z}$ (right) planes. The symbols are defined in Figure \ref{fig:apogee_abundance}.}
\label{fig:orbitparam}
\end{figure*}
%

Figure \ref{fig:segue_hist_metal} shows the MDF for the full sample (left panel) and different eccentricity intervals (middle and right panels) for the SEGUE stars from the overdensities and GSE. In the left panel, the MDF of GSE is similar to those from the different overdensities, with a slight difference toward the VMP regime. An equivalent result has been obtained by \cite{Naidu2021simulations}, with metallicities from the H3 survey.

In the middle panel, we show the MDFs for stars with $e > 0.7$, which is the range of the eccentricity characteristic of GSE. We estimate the medians and median absolute deviations of the distributions by bootstrap sampling with $10^{4}$ iterations, accounting for the uncertainties in metallicity. The GSE median [Fe/H] value in the SEGUE sample is $-1.39_{+0.01}^{-0.01}$ dex with median absolute deviation of 0.20 dex, which is in agreement with the median [Fe/H] of HAC-S, HAC-N, and VOD; $-1.51_{+0.06}^{-0.06}$, $-1.47_{+0.05}^{-0.05}$, and $-1.35_{+0.03}^{-0.03}$ dex, respectively. Differences in the median [Fe/H] between all overdensities and GSE are smaller than the typical uncertainties in SEGUE (${\geq}$0.1 dex). We also note that these listed median [Fe/H] values are well within literature determinations for GSE \citep[see discussion in][]{Limberg2022}.

In the right panel of Figure \ref{fig:segue_hist_metal}, we show the MDFs for $e < 0.7$. The most noticeable feature in this low-eccentricity regime is the excess of VMP stars in both HAC and VOD in comparison to their high-eccentricity counterparts (middle panel). This behavior is the same as previously identified in Figure \ref{fig:segue_ecc_hist}. The remaining stars from our local GSE selection (${<}1\%$) are also mostly VMP, which might indicate either contamination from other accreted substructures (\citealt{Helmi2020}) or a low-metallicity component of GSE itself, although we reinforce that the lack of low-eccentric GSE stars might be a bias due to our selection criteria.

\subsection{Elemental Abundances and Orbital Parameters}
\label{sec:apogee_analy}

We took advantage of the SEGUE sample to show that the bulk of the stars from the stellar overdensities have similar eccentricity and metallicity distributions, which are also compatible with GSE, suggesting that they share a common origin. We now explore in detail the chemical-abundance patterns of the overdensities with the available data from APOGEE. \par

Figure \ref{fig:apogee_abundance} shows the abundance of stars from HAC-S, HAC-N, VOD, and GSE in the [Fe/H]--[Mg/Fe], [Mg/Mn]--[Al/Fe], [Al/Fe]--[Fe/H], and [Ni/Fe]--[(C+N)/O] planes. In all panels, the yellow lines are isodensity contours associated with our GSE sample and the background 2D histograms show stars from APOGEE. All the overdensities clearly overlap with the contour of GSE and occupy the region of the chemical abundance planes dominated by accreted populations. This indicates that HAC-S/N and VOD share chemical properties compatible with GSE, favoring a scenario of a common origin for these substructures. However, we note that the origin of the VMP stars on low-eccentricity orbits identified in Section \ref{sec:segue_analy} could not be explored with APOGEE data given that it does not reach such low-metallicity regime. \par

To further analyze this chemodynamical compatibility, we 
select stars from HAC and VOD that have the characteristics of ancient accreted populations in the [Mg/Mn]--[Al/Fe] space \citep{Hawkins2015, das2020} and eccentricity consistent with GSE members ($e > 0.7$). The stars that do not satisfy these criteria (region of [Mg/Mn]--[Al/Fe] diagram delimited by red lines; \citealt{Limberg2022}) are represented by the empty symbols in Figure \ref{fig:apogee_abundance}, and we note that only a small fraction ($\sim$12$\%$) occupy the \textit{in situ} locus. This exercise makes it clear that the bulk of stars of the stellar overdensities studied in this work occupy the same region of GSE population in all chemical-abundance planes. \par

In Figure \ref{fig:orbitparam}, we show orbital parameters of HAC-S, HAC-N, VOD, and GSE. In the left panel, we see that the stellar overdensities occupy a well-defined region, which is dominated by GSE stars, in the total orbital energy ($E$) versus $L_z$ plane. The right panel ($e$--$L_{z}$) shows that the majority of HAC/VOD stars in this sample are on highly eccentric orbits and low average $L_z$, which are dynamical properties of GSE. In general, all the aforementioned chemodynamical properties suggest that the halo overdensities studied in this work share a common origin with GSE. \par
The stellar overdensities of our study are dynamically compatible with the GSE, which dominates the stellar content of the inner halo \citep{Deason2018, naidu2020}. This combination of orbital parameters was also identified in other works \citep{Simion2019,Balbinot2021} and is in contradiction with the hypothesis that these structures are dominated by stars on polar orbits or connected in a polar ring as speculated by \citet{Juric2008}. \par

\section{Conclusions and Final Remarks} %
\label{sec:conc}

In this Letter, we combined SEGUE, APOGEE, and Gaia data, together with \texttt{StarHorse} distances, to investigate the origin of both HAC and VOD and their potential link with GSE. With this goal, we performed a detailed chemodynamical analysis of these overdensities and compared them with GSE. Our main results are summarized below.

\begin{itemize}

    \item The stellar overdensities studied in this work show similar eccentricity distributions between them. VOD and HAC are composed mostly of stars with $e>0.7$ ($68\%$ and $65\%$, respectively), which is characteristic of GSE.
    
    \item HAC-S, HAC-N, VOD, and GSE have similar MDFs, within uncertainties, with median [Fe/H] values of $-1.52_{+0.05}^{-0.05}$, $-1.47_{+0.04}^{-0.05}$, $-1.37_{+0.04}^{-0.03}$, and $-1.40_{+0.01}^{-0.01}$ dex, respectively.
    
    \item The majority of stars from HAC-S, HAC-N, and VOD share a common region in the [Mg/Fe]--[Fe/H], [Mg/Mn]--[Al/Fe], [Al/Fe]--[Fe/H], and [Ni/Fe]--[(C+N)/O] planes. Furthermore, they overlap the GSE population in all these chemical-abundance spaces.

    \item We identified that HAC-S, HAC-N, and VOD exhibit contributions of metal-poor stars ($\rm[Fe/H] \lesssim -1.5$) on low-eccentricity orbits ($e \sim 0.5$). We speculate some scenarios for the origin of these stars as either a less eccentric and metal-poor portion of GSE or the contribution of other accretion event(s). We do not rule out the contribution/contamination of \textit{in situ} stars in the stellar overdensities, although this is difficult to reconcile with their low [Fe/H] values.
\end{itemize}

Our main result is that the majority of stars from HAC and VOD have chemodynamical properties characteristic of accreted populations, which are similar between them and indistinguishable from GSE. This serves as a constraint to models that simulate this accretion event and should reproduce stellar overdensities such as HAC and VOD in the inner Galactic halo. Furthermore, additional high-resolution spectroscopic observations of HAC and VOD members with different eccentricities will allow us to understand the origin of the group of VMP stars on low-eccentricity orbits.

\newpage

\acknowledgments
We thank the anonymous referee for useful comments that helped to improve this work.
H.D.P. thanks FAPESP proc. 2018/21250-9. G.L. acknowledges FAPESP (Proc. 2021/10429-0). J.A. acknowledges funding from the European Research Council (ERC) under the European Union’s Horizon 2020 research and innovation programme (grant agreement No. 852839). S.R. would like to acknowledge support from FAPESP (Proc. 2015/50374-0 and 2014/18100-4), CAPES, and CNPq. R.M.S. acknowledges CNPq (Proc. 30667/2020-7). A.P.V. acknowledges the DGAPA-PAPIIT grant IA103122. H.D.P., G.L., J.A, S.R., and R.M.S. thanks the ``Brazilian Milky Way group meeting´´, specially Helio J. Rocha-Pinto and Leandro Beraldo e Silva for the weekly discussions that based the development of this Letter. 

This research has been conducted despite the ongoing dismantling of the Brazilian scientific system.

\bibliographystyle{aasjournal}


\bibliography{bibliography.bib}




\end{document}